# Magnetic Mn₅Ge₃ nanocrystals embedded in crystalline Ge: a magnet/semiconductor hybrid synthesized by ion implantation

Shengqiang Zhou [1*], Wenxu Zhang [2], A. Shalimov [1], Yutian Wang [1, 4], Zhisuo Huang [2], D. Buerger [1], A. Mücklich [1], Wanli Zhang [2], H. Schmidt [3], M. Helm [1, 4]

[1] Institute of Ion Beam Physics and Materials Research, Helmholtz-Zentrum Dresden-Rossendorf, P. O. Box 510119, Dresden 01314, Germany

[2] State Key Laboratory of Electronic Thin Films and Integrated Devices, University of Electronic Science and Technology of China, Chengdu 610054, China

[3] Fakultät Elektrotechnik und Informationstechnik, Materialsysteme der Nanoelektronik, Technische Universität Chemnitz, 09107 Chemnitz, Germany

[4] Technische Universitat Dresden, 01062 Dresden, Germany

Corresponding author: Shengqiang Zhou, s.zhou@hzdr.de

# Abstract:

The integration of ferromagnetic $Mn_5Ge_3$ with the Ge matrix is promising for spin injection in a silicon-compatible geometry. In this paper, we report the preparation of magnetic $Mn_5Ge_3$ nanocrystals embedded inside the Ge matrix by Mn ions implantation at elevated temperature. By X-ray diffraction and transmission electron microscopy, we observe crystalline Mn5Ge3 with variable size depending on the Mn ion fluence. The electronic structure of Mn in $Mn_5Ge_3$ nanocrystals is $3d^6$ configuration, the same as in bulk $Mn_5Ge_3$. A large positive magnetoresistance has been observed at low temperatures. It can be explained by the conductivity inhomogeneity in the magnetic/semiconductor hybrid system.

# Background

Due to its compatibility to Si technology, Ge has attracted special attention as a host semiconductor for diluted magnetic impurity atoms. However, due to the low solid solubility of transition metals in Ge intermetallic compounds (mainly $Mn_5Ge_3$) tend to form in the Ge host[1-6]. $Mn_5Ge_3$ is a half-metallic ferromagnet with large spin polarization[7]. By first principles calculation, large spin-injection efficiency is expected by the integration of $Mn_5Ge_3$ within the Ge matrix[7]. Electrical spin injection and detection in Ge have been experimentally demonstrated [8,9]. Therefore, considerable work has been done to fabricate epitaxial $Mn_5Ge_3$ films as well as nanostructures [10-12]. The Curie temperature ($T_C$) of $Mn_5Ge_3$ is 296 K, which can be effectively increased by carbon doping. Spiesser *et al*. reported the epitaxial growth of $Mn_5Ge_3C_x$ films on Ge(111) [13]. When x is around 0.6, $T_C$ can be as high as 430 K. On another hand, some unknown nanoscale Mn-rich phases also form under particular conditions during MBE growth [14-19]. Those nanostructures can have $T_C$ much higher than 300 K. Besides MBE, ion implantation has been used to prepare ferromagnetic semiconductors as well as hybrids of ferromagnets embedded in semiconductors [20-24]. The advantages of ion implantation include the compatibility with conventional Si-chip technology and the lateral patterning. Patterning by ion implantation allows the synthesis of magnetic structures comprised of different magnetic phases. By carbon implantation into $Mn_5Ge_3$ and $Mn_5Si_3$, Sürgers *et al*. obtained lateral magnetic hybrid structures in the micrometer and submicrometer range [25]. In this contribution, we report the preparation of magnetic $Mn_5Ge_3$ nanocrystals embedded inside the Ge matrix by Mn ions implantation at an elevated temperature. We identify the formation of nanocrystalline $Mn_5Ge_3$ by X-ray diffraction and transmission electron microscopy. The magnetic, electronic and mangneto transport properties will be reported for this magnetic/semiconductor hybrid system.

# Methods

Nearly intrinsic Ge(001) wafers (n-type with the electron concentration of $10^{13}$-$10^{14}$ $cm^{-3}$) were implanted with 100 keV Mn ions at 673 K to avoid amorphization. It is worthy to note that we also used p-type Ge(001) as the substrates and got similar structural and magnetic properties. We varied the ion fluence to get samples with a large range of Mn concentrations, resulting in different structural and magnetic properties. The corresponding preparation and characterization parameters are listed in Tab I. Structural analysis was performed by synchrotron radiation X-ray diffraction (SR-XRD) at the Rossendorf beamline (BM20) at the ESRF with an X-ray wavelength of 0.154 nm. Magnetic properties were analyzed using a superconducting quantum interference device (SQUID) magnetometer (Quantum Design MPMS) with the field along the sample surface. X-ray absorption spectroscopy (XAS) measurements were performed at the beamline UE46/PGM-1 at BESSY II (Helmholtz-Zentrum Berlin). Magnetotransport properties were measured using van der Pauw geometry with a magnetic field applied perpendicular to the film plane. Fields up to 9 T were applied over a wide temperature range from 5 K to 300 K.

## Results and Discussion

(1) $Mn_5Ge_3$ nanocrystal formation

The SR-XRD 2θ- θ scans confirm the formation of $Mn_5Ge_3$ nanomagnets. As shown in Fig. 1, beside the main peaks from Ge(004) the diffraction peaks of $Mn_5Ge_3$(111), (002), (310), (222) and (004) are clearly visible. Note that compared to the work by Ottaviano *et al.*[20], SR-XRD reveals more $Mn_5Ge_3$ peaks even for a much smaller Mn-ion fluence due to the large flux of x-rays from the synchrotron source, which allows for the detection of small $Mn_5Ge_3$ nanocrystals. Therefore, we have to revisit the work by Ottaviano *et al*. They concluded that $Mn_5Ge_3$ nanocrystals formed by ion implantation are preferentially (002) oriented in the Ge(001) matrix[20]. However, Zeng *et al.*[1] prepared $Mn_5Ge_3$ layers by molecular beam epitaxy and they found the crystalline orientation as $Mn_5Ge_3$(001)//Ge(111). SR-XRD observations therefore lead us to conclude that $Mn_5Ge_3$ nanocrystals formed by Mn implantation are indeed randomly oriented inside the Ge(001) matrix, which is also supported by the magnetic properties shown later. We found nearly isotropic hysteresis loops with magnetic field along different directions. Note that in the work Jain *et al*. [26], the $Mn_5Ge_3$ nanocrystals were grown by annealing Ge:Mn films on Ge(001) substrates prepared by MBE. Most of $Mn_5Ge_3$ clusters (97%) have their *c*-axis perpendicular to the film plane. The accumulated literature data suggests that the growth of $Mn_5Ge_3$ nanocrystals from the Ge matrix is different from the $Mn_5Ge_3$ thin films.

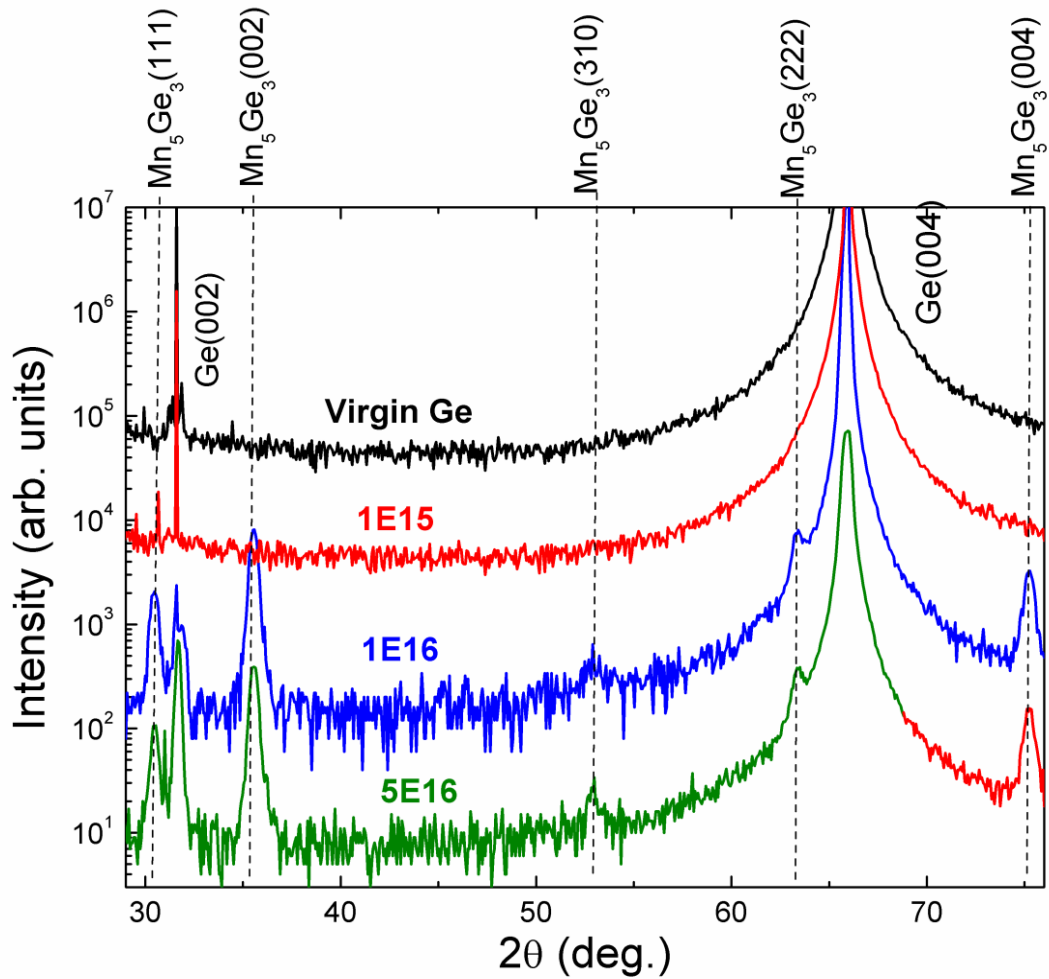

**Fig. 1 XRD 2θ-θ scans revealing the formation of $Mn_5Ge_3$ nanomagnets. Beside the main peaks from Ge(004), the diffraction peaks of $Mn_5Ge_3$(111), (002), (310), (222) and (004) are clearly visible**

Figure 2 shows cross-section TEM images of samples 1E16 and 5E16. The white contrast spots are from precipitates, which are located in the depth between 20-120 nm, in agreement with the depth profile of Mn ion implantation. The average crystallite size is increased from 5 nm to 11 nm with increasing Mn fluences from $1\times10^{16}$ to $5\times10^{16}$ cm$^{-2}$. For detailed analysis we focus on the sample 5E16. As shown in Fig. 2(b), the well-defined Moiré patterns are a strong indication for monocrystalline precipitates embedded in a crystalline matrix. Using high resolution TEM the precipitates can be identified to be $Mn_5Ge_3$ as shown in Fig. 2(c) and (d). Figure 2(d) is the Fast Fourier Transform (FFT) of the image shown in Fig. 2(c). FFT reveals lattice spacings amounting to 0.298 nm (indicated by the open circles) and 0.623 nm (indicated by the open squares), which correspond to $Mn_5Ge_3$(111) and (001), respectively.

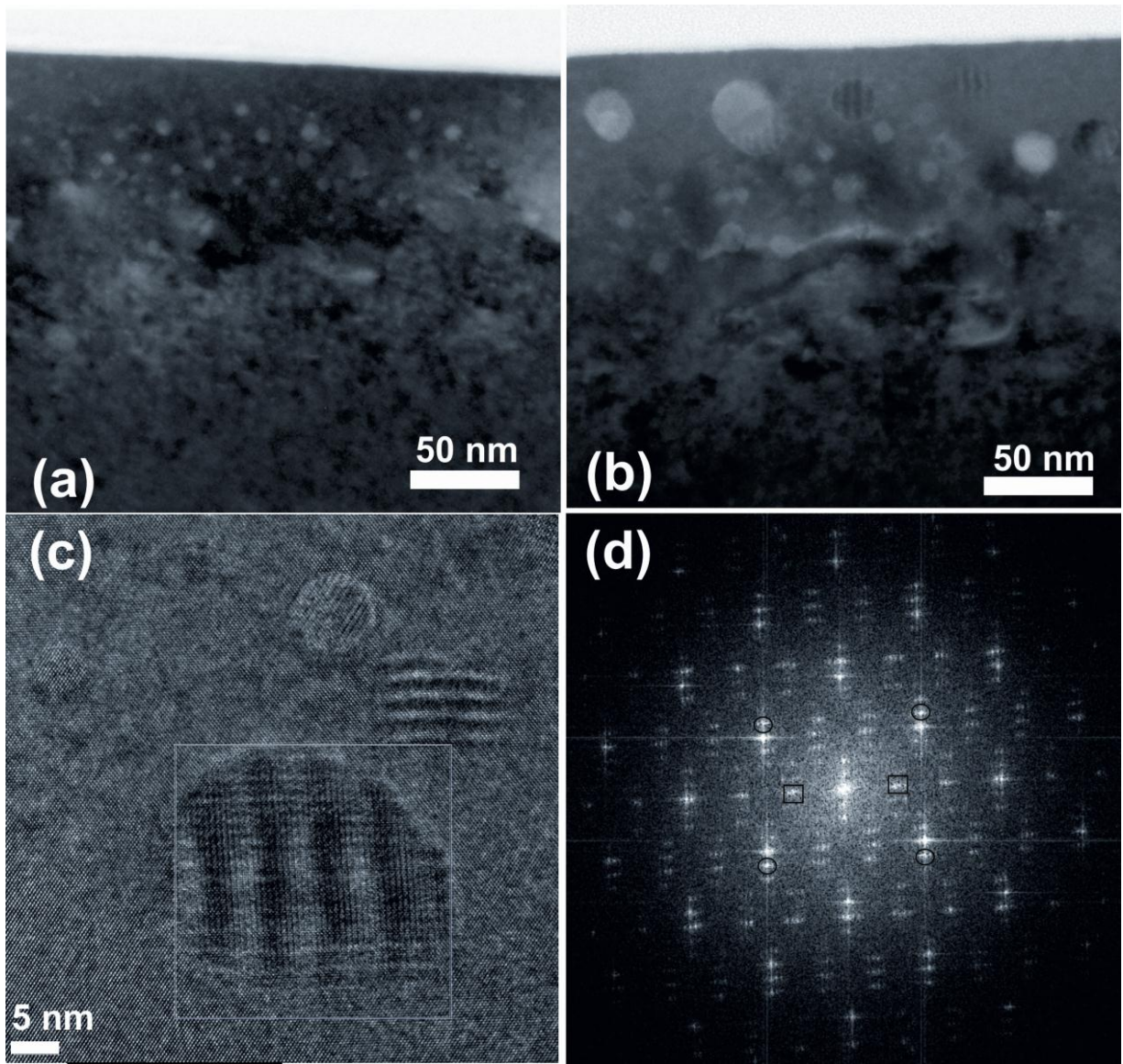

**Fig. 2 Transmission electron microscopy (TEM) image of the cross-section showing the formation of precipitates (a) 1E16, (b) 5E16 and high resolution TEM for an individual $Mn_5Ge_3$ particle (c) in sample 5E16. (d) FFT of the precipitate indicated in (c).**

(2) Magnetic properties

Figure 3(a) shows the zero field cooled and field cooled (ZFC/FC) magnetization curves in a 50 Oe field for different Mn fluences. The FC curve for sample 1E15 completely overlaps with the corresponding ZFC curve at around zero. Magnetic $Mn_5Ge_3$ nanocrystals can be excluded in this sample, in consistence with the SR-XRD observation, except they are very small and dilute beyond the detection limit of SR-XRD. For samples 1E16 and 5E16, a distinct difference in ZFC/FC curves was observed. The ZFC curves show a gradual increase at low temperatures peaking at different temperatures, while FC curves monotonically increase with decreasing temperature. The width of the peaks in the ZFC curves is due to the size distribution of $Mn_5Ge_3$ nanocrystals, as shown in TEM images (Fig. 2). In this paper, we take the temperature (Tmax) at the maximum of the ZFC curve as the average blocking temperature, listed in Tab. 1.

Figure 3(b) shows the magnetization versus field reversal (M-H) of all samples measured at 5 K. Hysteretic behaviors were observed for samples 1E16 and 5E16. With increasing Mn concentration, the saturation magnetization is increased from 10.1 emu/cm$^3$ to 69.2 emu/cm$^3$ (by assuming the implanted depth of 100 nm), and the coercivity is increased from 0.22 T to 0.26 T. At 300 K, sample 5E16 only shows field induced magnetization (see the inset of Fig. 3(b)). The saturation magnetization of sample and 5E16 (1E16) is 69.2 (10.1) emu/cm$^3$, corresponding to around 1.5 (1.1) $\mu_B$/Mn, which is smaller than 2.6±0.5 $\mu_B$/Mn reported in Ref. 1. That means not all of the implanted Mn ions form the ferromagnetic $Mn_5Ge_3$ phase.

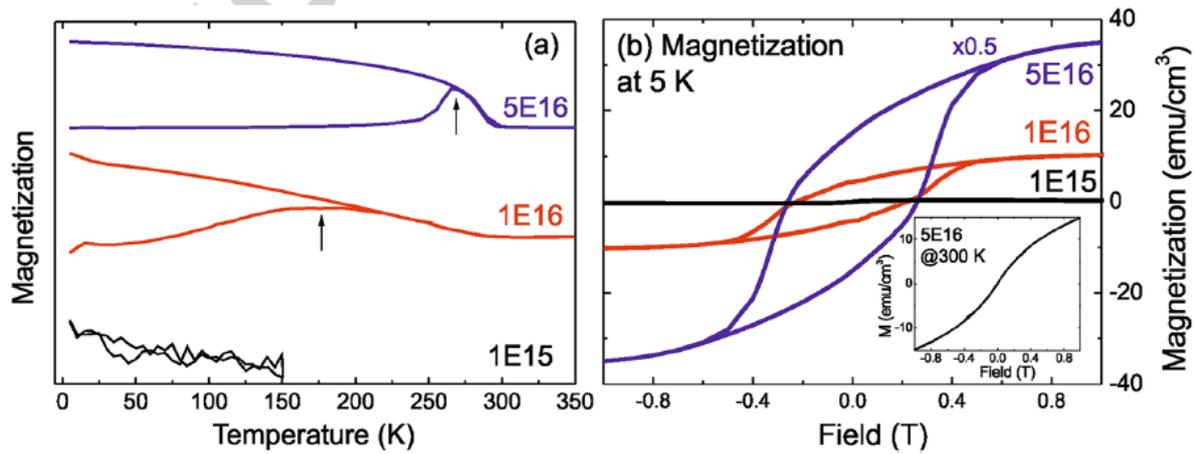

**Fig. 3 (a) Magnetization curves with an applied field of 50 Oe after ZFC/FC for the Mn implanted Ge. The lower branches are ZFC curves, while the upper branches are FC curves. With increasing fluence, the $Mn_5Ge_3$ nanocrystals are growing in size, resulting in a higher blocking temperature. The curves are vertically shifted to increase the visibility. (b) Hysteresis loops measured at 5 K for Mn implanted Ge with different fluence and the inset shows the magnetization at 300 K for sample 5E16.**

We also compared the magnetization between the in-plane and out-of-plane directions at 5 K for sample 5E16 (not shown). In contrast to Refs. 1 and 8, there is no detectable magnetic anisotropy. For bulk $Mn_5Ge_3$, the magnetic easy axis is [001]. The absence of magnetic anisotropy in our samples is due to the random crystallographic orientation of the $Mn_5Ge_3$ nanocrystals.

As shown in Fig. 3(b), the hysteresis loop is not square-like. The distribution of coercivity field is due to the size distribution of the nanomagnets as evidenced by TEM images and is also possibly due to the random distribution of nanomagnets easy axis. According to the Stoner and Wohlfarth model for single-domain magnetic nanoparticles the maximum coercive field gives the anisotropy field $\mu_0 H_{a2} = 0.26$ T for sample 5E16. Using the bulk saturation magnetization ($M_S$) for $Mn_5Ge_3$ (1100 kA/m) [26], one can deduce the anisotropy constant: $K_2 = \mu_0 H_{a2} M_S / 2 \approx 1.4 \times 10^5$ J/m$^3$, which is smaller than the value reported by Jain *et al.* [26] Based on the Néel–Brown model, the volume for a nanomagnet $V = 25 k_B T_{max} / K_2$ ($k_B$ as the Boltzmann constant) we calculate the average diameter of $Ge_3Mn_5$ clusters in sample 5E16 to be ~ 10.8 nm ($T_{max} = 270$ K). The average diameter is in good agreement with the results obtained by TEM. However, the average diameter for sample 1E16 deduced from the ZFC magnetization is as large as 9.5 nm, which is much larger than the value from TEM observation.

The magnetic properties of the Mn implanted Ge were also investigated by X-ray magnetic circular dichroism (XMCD) at Mn $L_{2,3}$ edge. Right before XAS measurements, the sample was etched in deionized water for 2 minutes to remove the surface oxide layer [27]. Fig. 4(a) presents the Mn $L_{2,3}$ XAS measured in total electron yield (TEY) mode at around 4.5 K. μ+ and μ− represent the absorption intensity with the direction of magnetization parallel and antiparallel to the photon helicity, respectively. As shown in Fig. 4(a), after etching we obtained very similar spectra as was reported for ferromagnetic $Mn_5Ge_3$ [28]. The XAS spectra can be classified into the $2p_{3/2}$ (~ 641 eV) and $2p_{1/2}$ (~ 651 eV) absorption regions. The shape of the main feature indicates the itinerant nature of ferromagnetic $Mn_5Ge_3$. On the other hand, the weak shoulders appear at 642 and 644 eV and the doublet structure is observed in the $2p_{1/2}$ excitation region, which could be related with some diluted Mn impurities in the Ge matrix [21] or oxidized Mn [29]. Fig. 4(b) shows the XMCD spectrum, revealing a large negative signal (~ 641 eV) and a small positive signal (~ 644 eV) in the $2p_{3/2}$ region and a larger positive signal (~ 651.5 eV) in the $2p_{1/2}$ region. Note that the shoulders and the doublet in XAS spectra are hardly resolvable in the XMCD spectrum, which indicates that the oxidized Mn ions have no contribution to the ferromagnetism. According to the sum rule, the integrated intensity of the XMCD signal in the whole region is proportional to the orbital magnetic moment relative to the spin magnetic moment. In the present XMCD spectrum, the integration is nearly zero, indicating that the orbital magnetic moment is negligible for $Mn_5Ge_3$. Comparing our experimental results with the published calculations, the electronic structure of Mn ions can be assumed to be in the $3d^6$ configuration without spin-orbit interaction [28, 30].

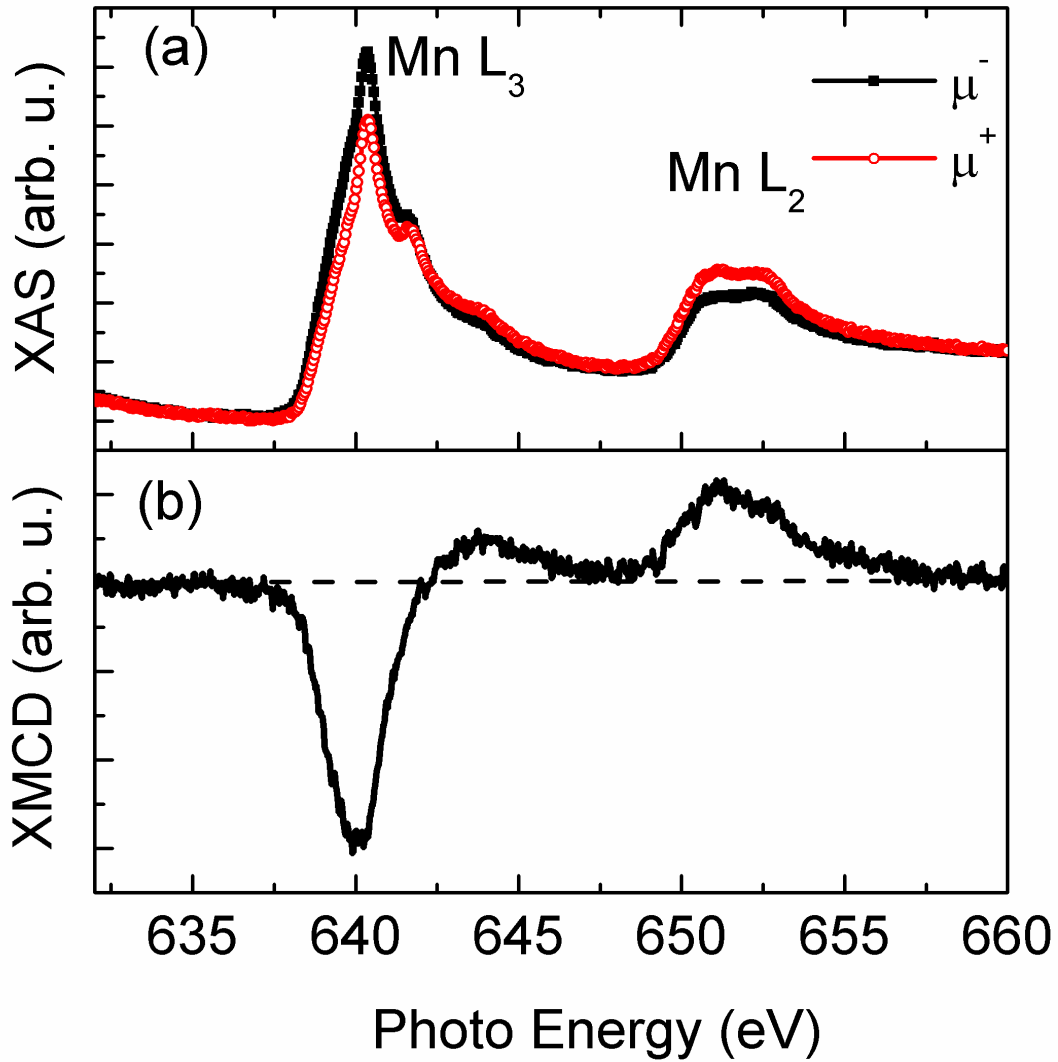

**Fig. 4** Mn L3, 2 total electron yield (TEY) (a) XAS for magnetization and helicity parallel (μ+) and antiparallel (μ-) and (b) XMCD (μ+ - μ-) measured at around 10 K under an external field of 6000 Oe applied perpendicular to the surface.

(3) Magneto-transport properties

All three samples show p-type conductivity and large magnetoresistance (MR) effect. Fig. 5(a) shows the measurement results for sample 5E16. The MR is defined as [R(H)-R(0)]/R(0), where R(H) is the sheet resistance at a field of H, R(0) is the sheet resistance at zero field. One can see that MR is positive and may saturate at a large field. Different from ferromagnetic semiconductors or metals, there is no hysteresis in MR curves for the $Mn_5Ge_3$/Ge hybrids. In this case the spin scattering should have very small contribution to transport. Similar positive MR effect has been reported also for GeMn-nanocolumns/Ge hybrids [15]. Note that the Ga-doped Ge shows only neglectable MR effect as shown in Fig. 5(a). The MR effect can be interpreted by the inhomogeneity of the sample: the different conductivity and Hall resistivity of GeMn and Ge (or Mn-rich and -poor regions). We modelled the hybrid system where Mn-rich nanoparticles were embedded in the Ge matrix by a 2D slice as in the work of Yu *et al.* [31]. Under steady-state conditions, the continuity of the current requires that $\nabla \cdot [\sigma \cdot \nabla U(x, y)] = 0$, where $U(x, y)$ is the electrostatic potential at position (*x*, *y*) on the 2D slice. The materials were fully characterized by their conduction matrices, which vary with the position of different materials. Thus, the finite element method (FEM) proposed by Moussa *et al.* [32] was used. We applied constant potential between two electrodes and calculated the induced averaged potential difference at the other two electrodes in the geometry of van der Pauw method. The current normal to the boundary of the slice was set to zero (the natural boundary condition). The transport properties of the matrix and nanocrystal are simple characterized by the conductivity matrix with the components:

$$\sigma_{xx}(\beta) = \sigma_{yy}(\beta) = \sigma(0)/[1+\beta^2]$$

$$\sigma_{xy}(\beta) = -\sigma_{yx}(\beta) = -\sigma(0)\beta/[1+\beta^2]$$

Where σ(0) is the zero-field conductivity and $\beta = R_H \sigma(0) \mu_0 H$ in which $R_H$ is the Hall coefficient of the materials and $\mu_0$ is the susceptibility in vacuum.

The material parameters of the matrix are chosen to be those of Ge: $\sigma_H^{Ge}(0) = 10^4$ $\Omega^{-1} \cdot m^{-1}$, $R_H^{Ge} = 10^{-6} C^{-1} \cdot m^3$. The material parameters of the nanocrystal are: $\sigma_H^{GeMn}(0) = a\sigma_H^{Ge}(0)$ and $R_H^{GeMn} = bR_H^{Ge}$. There are two free parameters *a* and *b* which are the ratios of conductivity and Hall coefficient of the two phases, respectively. Both the conductivity and the Hall coefficient are functions of temperature. The resistance of the system is calculated by finite element methods (FEM) where a constant current is applied and the corresponding voltages are measured in the geometry of van der Pauw method. The calculated curves are presented in Fig. 5(b). The experimental MR curves can be well reproduced by FEM calculations. The *a* and *b* values used in FEM calculations are shown in Fig. 5(c). The MR magnitude is sensitive to the ratio of conductivity of the two constitutes. Beside the magnetoresistance, the samples also show anomalous Hall resistance (*i.e.* the Hall resistance deviates from a linear behaviour), which can be explained by two kinds of carriers with different mobilities [33]. On the other hand, we have to note the rather large discrepancy in the MR magnitude between the experimental and modelled values. In the model, for simplifying we neglect the anomalous Hall effect in the GeMn constitute, which may induce this discrepancy. Also, in order to account non-monotonic dependence of MR on temperature [see Fig. 5(c)], we have to vary parameters *a* and *b* accordingly. The decrease of a and b at temperature below 50 K cannot be understood and is the aim for the future work.

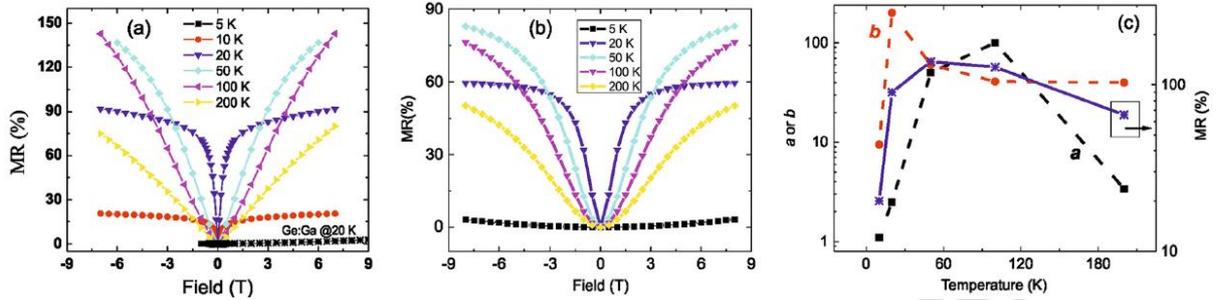

**Fig. 5 (a) Magnetoresistance (MR) for sample 5E16 measured at different temperatures and the result for Ga implanted Ge (symbol \*) is shown for comparison, (b) The calculated MR by considering two conductivity components, (c) parameters a and b used in the FEM calculations as well as the MR data at 6 T at different temperatures .**

# Conclusions

We have prepared magnetic $Mn_5Ge_3$ nanocrystals embedded inside the Ge matrix by Mn ion implantation into Ge substrates. The crystalline size of $Mn_5Ge_3$ can be tuned by varying Mn fluence. The Mn ions in $Mn_5Ge_3$ nanocrystals are in $3d^6$ configuration. Large positive magnetoresistance has been observed in the $Mn_5Ge_3$/Ge hybrid system. It could be due to the inhomogeneity in samples with constitutes having different transport properties.

# Competing interests

The authors declare that they have no competing interests.

# Authors'contributions

SZ designed the experiments and wrote the manuscript. WZ and ZW made fittings for the magnetoresistance data. AS performed the XRD measurement. YW carried out the XMCD and XAS measurements. DB and HS helped during magneto-transport measurement. AM performed the TEM characterization. WZ supervised the fitting of the magnetoresistance data. MH supervised the whole work. All authors have read the manuscripts.

# Acknowledgements

The work was supported by the Helmholtz-Gemeinschaft Deutscher Forschungszentren (HGF-VH-NG-713) and by the International Science & Technology Cooperation Program of China (2012DFA51430). The author (HS) thanks the financial support under DFG SCHM1663/4-1.

**Table I: Sample identification, structural and magnetic parameters.**

| Sample identifier | Mn fluence | Concentration | $Mn_5Ge_3$ (XRD) | $T_{max}$ (ZFC/FC) | Average Diameter |
|---|---|---|---|---|---|
| 1E15 | $1\times10^{15}$ cm$^{-2}$ | 0.2% | / | / | / |
| 1E16 | $1\times10^{16}$ cm$^{-2}$ | 2% | Yes | 185 K | 5 nm |
| 5E16 | $5\times10^{16}$ cm$^{-2}$ | 10% | Yes | 270 K | 11 nm |